\documentclass[aps,pre,showpacs,preprintnumbers,amsmath,amssymb,superscriptaddress]{revtex4}

\usepackage{graphicx}
\usepackage{bm}
\begin{document}
\title{Statistics of Jamming in the discharge of a 2-D Silo}
\author{Gabriel P\'erez$^*$}
\affiliation{Departamento de F\'{\i}sica Aplicada, Centro de Investigaci\'on 
y de Estudios Avanzados \\
del Instituto Polit\'ecnico Nacional, Unidad M\'erida \\
Apartado Postal 73 ``Cordemex'', 97310 M\'erida, Yucat\'an, M\'exico}
%(Started June 21, 2006)
\date{\today}
\begin{abstract}
Jamming and avalanche statistics are studied in a simulation of 
the discharge of a polydisperse ensemble of disks from a 
2-D silo. Exponential distributions are found 
for the avalanche sizes for all sizes of the exit opening,
in agreement with reported experiments. The average avalanche size 
grows quite fast with the size of the exit opening. Data 
for this growth agree better with a critical divergence with a large
critical exponent, as reported for 3-D experiments, than with 
the exponential growth reported for 2-D experiments.
\end{abstract}
\maketitle

\section{introduction}

The discharge of granular matter from a container is one of the 
most common phenomena of everyday life, but it is not yet 
completely understood. Be it the pouring of salt from a salt-shaker, 
of corn form a hopper, or of gravel from a truck, we count only with
a few empirical rules to explain the process \cite{beverloo1,nedderman1}.
Due to both its intrinsic interest and its obvious practical applications 
this phenomenon has received the attention of many researchers in
recent times. It also provides a paradigmatic example of 
\emph{jamming} \cite{liu1}, a recently proposed transition to a 
particular state of matter (also called \emph{fragile matter})
that can support some stresses, called \emph{compatible\/}, 
but flow under \emph{incompatible\/} ones. Jammed systems are then non-solids
that, under the influence of external forcing, develop some particular
structures that block their flow \cite{zuriguel1}.

A particularly simple and common example of granular matter discharge
is given in hoppers and silos \cite{hopper-silo}. 
Three states of flow have been 
identified in these systems: dilute (gas-like), dense (liquid-like)
and jammed (static, solid-like) \cite{hou1,zhong1}. The transition 
from dilute to dense, as a function of the exit opening,
is discontinuous, and shows hysteresis \cite{hou1}.
In these containers jamming is known to appear as soon as the size 
of the exit hole is reduced to a few times the
average diameter of the particles inside. Experimental work in both 
2-D and 3-D hoppers \cite{to1,to2} and silos 
\cite{to2,zuriguel1,zuriguel2,hou1,zhong1}
have shown that jamming depends only on the ratio between
particle and exit hole sizes, as long as the diameter and height of the
silo are large enough, that is, the silo is in its ``thermodynamic limit''.

Among the many questions still remaining in the dynamics of
silos, one of the simplest and more fundamental is is the possible
existence of a critical hole size such that for larger holes the
flow cannot jam. The existence of such a size ---actually, of a critical
value for the ratio $R = \mbox{exit-hole-size/grain-size}$--- has been shown in
experiments for several types of granular media, including smooth and
rough spheres, rice grains and lentils \cite{zuriguel2}. This is
a somewhat puzzling result, since the same work shows that the 
distribution of avalanche sizes can be very well fitted to 
an exponential, and this in turn is consistent with a simple model where 
each grain ---maybe cluster of grains--- has a given probability
of exiting the silo, uncorrelated to the behavior of other grains 
(clusters). 
Accepting this model, it becomes difficult to understand how
a fully uncorrelated process can give rise to 
the long distance correlations one usually associates to criticality. It
should also be mentioned that the experiments carried in 
\cite{to1,to2} also support the hypothesis of a fixed probability of
exiting the silo for each grain, but point towards a 
probability of jamming that is exponentially decaying on $R$; however, 
Ref.~\cite{to2} also argues that it may be possible to fit its
experimental data to either one of the two behaviors.

In this work we have carried out a simulation of a 2-D silo with
variable hole size, with the intention of getting some information
on the statistics of its discharge, and on the possible presence of
criticality in this process. Even though simulational approaches 
cannot reproduce completely the dynamics of real experiments, they do give
very good approximations to real flows, and should be able to
find signatures of critical behavior, if present. They have the
added advantage of allowing for continuous and unconstrained adjustments 
in the main parameters of the flow. There are some obvious 
limitations to this particular simulation that puts it at a 
disadvantage with respect to actual experiments, chiefly the fact that 
tracking the very long avalanches that appear for large exit hole  
sizes consume an inordinate amount of computer time. Still, a 
systematic study of the avalanches for different exit hole sizes 
allows the identification of a well defined trend.

\section{Simulation} 

The simulations reported here were done over an ensemble
of $N$ polydisperse disks, with diameter given by
$d_i = d_{ave} + x \Delta d$, where $d_{ave}$ 
is the average diameter, $\Delta d$ is its maximum
fluctuation, and $x$ is randomly
chosen from the $[-1,1]$ uniform distribution. 
The silo has a bottom size $D$ and indefinite height. At the 
center of the bottom there is a hole of size $d_H$. The disks
have a 2-D 
mass density $\sigma$, and the gravitational acceleration $g$ acts
in the negative $z$ direction. 
Upon contact, the disks interact with the (perfectly rigid) walls of
the silo, and with each other, via a linear spring with a constant 
that on loading has the value $\kappa$, and on unloading is reduced by 
a restitution factor $\epsilon$. This is an implementation of 
the \emph{linear spring-dashpot model} \cite{cundall1,schaefer1},
using the two-couplings approach given in \cite{walton1}. 
This approach is commonly used because of its robustness and simplicity.  
The interaction is complemented by dynamic and static friction, 
using for both the same 
coefficient $\mu$. For the static part the the tangential spring model 
given in \cite{cundall1} is used, following the specific formulation 
of \cite{schaefer1}, with the corrections given in \cite{brendel1}. 
Equations of motion were 
integrated using a velocity-Verlet algorithm \cite{allen1}. 

The only fixed numerical input in the problem is given by 
the gravitational constant; all quantities can be
scaled, say to natural units where the disks' average diameter 
and mass are set to one. In this work we have preferred to implement a 
numerical experiment with standard units, and have used $N = 2000$
disks with 
$d_{ave} = 0.5$, $\Delta d = 0.05$, $D  = 15$, $\sigma = 0.8$ and 
$\kappa = 4\times10^6$, with all quantities given in the cgs system. 
This value for $\kappa$ is not as large as could be expected for 
some hard real systems (steel or glass spheres, for instance), but it 
allows for a more efficient use of computer time. It can be 
realistic enough for 
softer grains, like the rice or lentils used in \cite{zuriguel2}. 
It has also been reported that changes of up one or two orders of 
magnitude in the stiffness of the grains have little 
influence on the results of these types of simulations \cite{silbert1}.
For the adimensional quantities $\mu$ and $\epsilon$ we have set 
$0.5$ and $0.9$ respectively. Gravity is fixed as $g = 981$.
The ratio between the diameters of the disks and the silo 
gives $D/d_{ave} = 30$, and for the number of disks used the silo
gets filled up to a height of around 2.5 times $D$. These two values are 
large enough to put the silo in the thermodynamic limit
\cite{zuriguel1,zuriguel2,hirshfeld1}, at least for the 
3-D case. A time-step of 0.01 times the disk-disk collision time 
$t_{coll}$, which for the linear interaction used here is given by
$$
t_{coll} = \pi \sqrt{\frac{m}{\kappa}},
$$
neglecting a weak dependence on $\epsilon$. For later convenience, 
we also define the time and speed scales $t_0 = \sqrt{d_{ave}/g}$
and $v_0 = g\, t_0$.

The simulation proceeds as follows: first the silo is closed from below
and the disks are placed in a regular grid with random initial velocities.
The system is then allowed to relax, under the influence of
gravity, up to the moment where the maximum
speed detected is some small fraction of $v_0$. At his point a hole of
length $d_h$ appears at the center of the bottom line, and disks start 
pouring out of the silo. These falling disks are followed until 
their centers are a distance $\approx 1.3\, d_{ave}$ below the bottom,
and at this moment they are eliminated from the exiting flow and
re-injected on top of the system, at a distance of $5.5\, d_{ave}$ from 
the surface of the dense aggregate of disks. In re-injection
the disks preserve their $z$ velocity, but their horizontal velocity 
is set to zero. The re-injection point is chosen so as to keep the
top of the material roughly flat. With these conditions the observed 
flow is of a mixed type, neither massic nor funnel-like; the disks close 
to the center of the silo fall faster than those close to the walls, 
but the speed difference is not too large.

Occasionally, an arch is formed above the exiting hole and the flow stops. 
Given that we are including static friction, these arcs are not always 
convex. This jammed state is detected by checking that both (1)
the \emph{maximum\/} speed in the system is less than $v_0/c$,
and
(2) no disk has exited the silo in a time longer than 
$c\, t_0$. It has been found that for the purposes of this simulation
$c = 8$ is adequate.
%Trial and error show that a factor of 8 is enough.
Once these two conditions are fulfilled, the silo receives a tap 
given by vertical displacement $z_{tap} = A \sin( 2 \pi \nu t)$, 
applied for half a period. Here we have used $A = 0.6$ and $\nu = 8.0$.
In most cases this tap is enough to break
the arch or arches that are blocking the flow; however, given that the 
tap moves the whole material in parallel, it does occasionally happen 
that there is not enough rearrangement of the disks as to break the 
blockage. It is possible therefore with this unjamming protocol
to get \emph{null avalanches}, which are time intervals between two 
taps where no material flows out of the silo. These null avalanches 
are highly correlated among themselves, in the sense that, for small 
openings, they tend to appear next to each other in the time record. 
This type of events have also appeared in the experiments reported 
by Zuriguel~\cite{zuriguel3}.
Null avalanches are common for very small hole sizes, less so for
larger openings. 

\section{results}

\begin{figure}
\includegraphics[width=14.0cm]{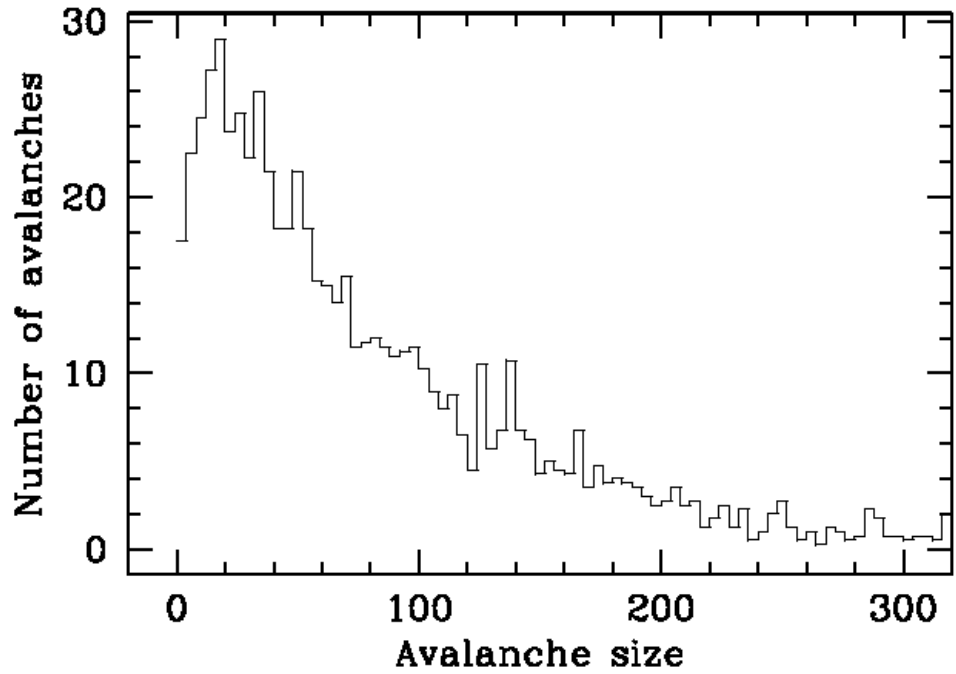}
\caption{Histogram of avalanches for
$R = 3.6$. Bin size has been set to $4$, and null avalanches have
not been included.}
\label{small-s}
\end{figure}

The simulation has been carried on for hole sizes from 
1.70 to 2.25, in steps of 0.05, corresponding to 
hole/particle ratios $R$ from 3.4 to 4.5 in steps
of 0.1. In all cases we have performed several 
runs starting from different grain configurations.
For each size we have obtained at least 1000 avalanches. 
For all hole/particle ratios the distributions of avalanches $n(s)$
show basically an exponential form,
except for a spike at $s = 0$ (null avalanches),
and a weak dip for small $s$ (see Fig.~(\ref{small-s})). 
These two characteristics are probably a peculiarity of the 
method used here to unjam the silo. It should be noticed
that the decrease in $n(s)$ found for small $s$ is not as 
pronounced as the one reported from the experiments~\cite{zuriguel2}.
As for the probability of finding null avalanches,
it goes from a maximum of 0.20 at $R = 3.4$ to a minimum of 0.019
at $R = 4.4$, but there is not enough statistics
as to be able to predict their presence or absence for 
larger values of $R$.

\begin{figure}
\includegraphics[width=14.0cm]{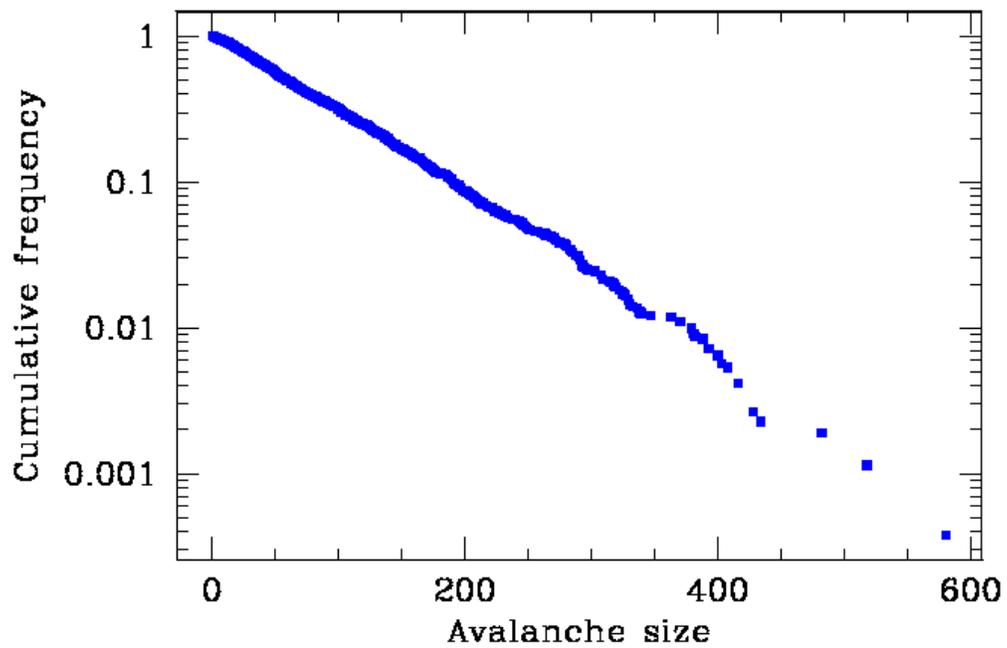}
\caption{Cumulative curve $w(s)$ for
$R = 3.6$. Null avalanches are excluded.}
\label{cumulative}
\end{figure}

\begin{figure}
\includegraphics[width=14.0cm]{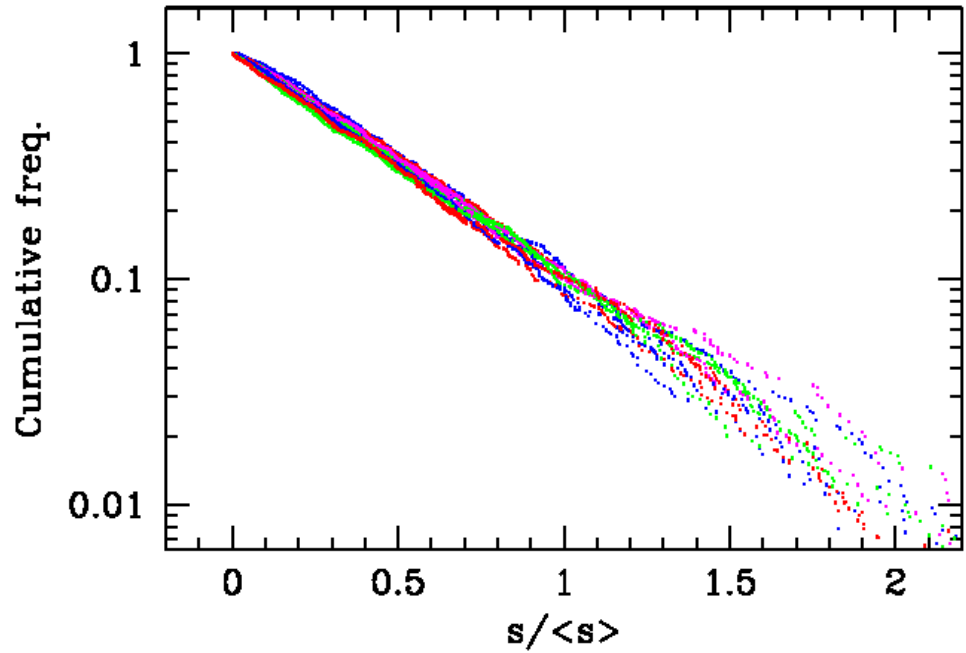}
\caption{Collapse of the cumulatives $w(s)$ for all values of $R$
considered. The regions of non-exponential behavior found
for small $s$ have some incidence over the observed dispersion, since
in all cases the normalization is $w(1) = 1$.}
\label{colapso}
\end{figure}

To avoid having to fix a bin size in the histograms, we 
have used the normalized cumulative distribution
$$
w(s) = {\sum_{s^\prime = s}^{\infty} n(s^\prime)}/
{\sum_{s^\prime = 1}^{\infty} n(s^\prime)},
$$
that is, we count the number of avalanches with $s$ or more 
disks. Notice that in this measure we are leaving out 
the null avalanches. For a properly normalized exponential 
distribution the normalized cumulative happens to be identical
to the distribution itself. Fig.~(\ref{cumulative})
shows the cumulative avalanche distribution for $R = 3.6$. 
Given the claim that the distribution is exponential for all
cases, it should be possible to scale $s$ to obtain a collapse
of all cumulatives. This is shown in Fig.~(\ref{colapso}).
Even so, it should be remembered that 
the distribution is not a perfect exponential, due to the 
smaller probabilities found for very small avalanches. This 
effect is almost imperceptible in the cumulatives.

The main question that remains to elucidate is the behavior 
of the average avalanche $\langle s \rangle$ with respect to 
the hole/particle ratio. As intuitively expected, one finds 
a rapidly growing curve. This growth can be interpreted as 
evidence of an exponential divergence of the form 
$\langle s(R) \rangle \approx s_0 \exp(R/R_0)$, meaning that 
there is a non-zero probability of jamming for any size of the 
exit hole, even if for large openings the typical avalanche 
becomes astronomically large. However, as pointed out in~\cite{to2},
it is also possible to obtain a good fit to a power law of the form 
$\langle s(R) \rangle \approx s_0/(R_c - R)^\gamma$. 

\begin{figure}
\includegraphics[width=14.0cm,angle=0]{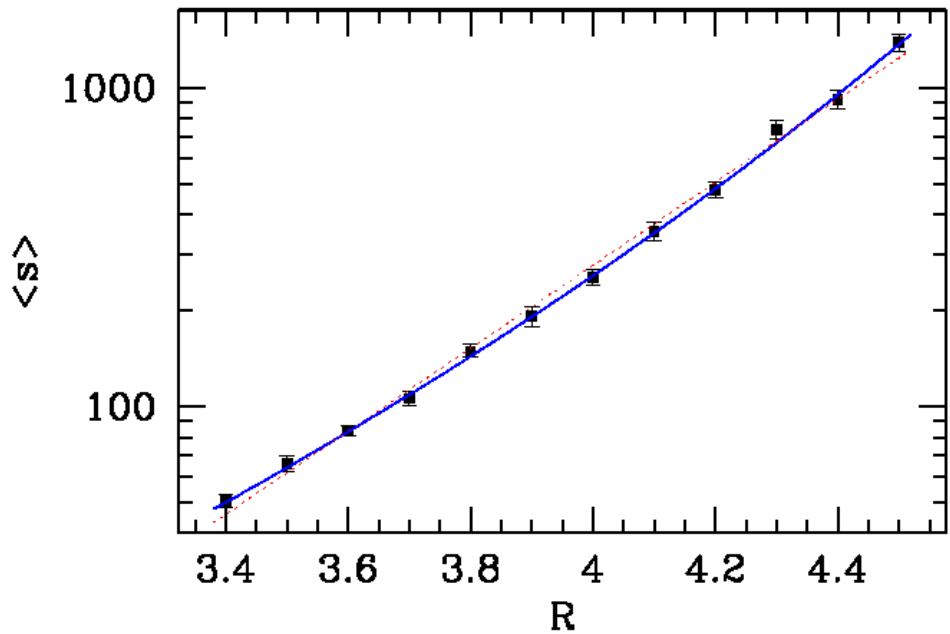}
\caption{Scaling of the average avalanche $\langle s \rangle$ against
the exit-hole/disk ratio $R$. The continuous line give the power-law
fit, with $R_c = 6.7$ and $\gamma = 8.16$. The dotted line shows the best
linear fit, which corresponds to an exponential growth for the averages.}
\label{powerlaw}
\end{figure}

Trying both types of fit for the results of this simulation, 
we find that the the avalanche averages 
have a better fit to the power-law expression, with 
$R_c = 6.7 \pm 0.4$ and $\gamma = 8.16 \pm 1.10$.
For this fit the $\chi^2/\mbox{dof}$ is $0.41$. For the exponential 
fit we get $R_0 = 0.768$ and a $\chi^2/\mbox{dof}$ of $1.61$.
The two different fits are shown in the semi-log graph given in 
Fig.~(\ref{powerlaw}). 

It is then clear that a power-law divergence is favored over
an exponential behavior; it is is also clear, however, that 
with the available data the difference between the two fits 
is not really large enough as to allow for a definite conclusion. 
Moreover, some other types of divergence have been hinted at, 
like an essential singularity given by 
$\langle s \rangle \approx s_0 \exp[1/(R_c - R)]$. Still, the results
agree with the most extensive experiments performed at this time,
and therefore adds support the existence of criticality in the 
jamming of silos. This leaves open the more fundamental question about
the origin of the correlations that may lead to critical behavior
in this type of phenomena.

\newpage



\begin{references}
\bibitem{hopper-silo} Here a silo is any hopper with a flat
bottom.
\end{references}
\end{document}